%% file: draft_dsnb_method.tex
\documentclass[aps,prd,reprint,groupedaddress,nofootinbib]{revtex4-1}

\usepackage{footnote}

\usepackage{graphicx}	
\usepackage[usenames]{color}
\usepackage{appendix}
\usepackage{booktabs}
\usepackage{amsmath}

\usepackage{hyperref}
\usepackage{natbib}
\hypersetup{
    colorlinks=true,
    linkcolor=blue,
    citecolor=blue,        
    filecolor=magenta,
    urlcolor=cyan,
}

\def\nue{\mathrel{{\nu_e}}}

\def\nux{\mathrel{{\nu_x}}}

\def\barnue{\mathrel{{\bar \nu}_e}}

\def\msun{\mathrel{{M_{\odot} }}}
\def\mco{\mathrel{{M_{\rm CO} }}}
\def\xic{\mathrel{{\xi_{2.5} }}}

\newcommand{\km}{KIHNN}
\newcommand{\df}{DSNB}
\newcommand{\bhfc}{BHFC}
\newcommand{\nsfc}{NSFC}
\newcommand{\n}{neutrino}

\newcommand{\sn}{supernova}

\newcommand{\be}{\begin{equation}}
\newcommand{\ee}{\end{equation}}
\newcommand{\ba}{\begin{eqnarray}}
\newcommand{\ea}{\end{eqnarray}}

\defcitealias{Vartanyan:2023zlb}{VB23}
\defcitealias{Horiuchi:2017qja}{H18}
\defcitealias{Horiuchi:2020jnc}{H21}
\defcitealias{Kinugawa:2014zha}{KIHNN}

\begin{document}

\title{The diffuse supernova neutrino background: an
update with modern population synthesis and core-collapse simulations}

\author{Cecilia Lunardini}
\email[]{Cecilia.Lunardini@asu.edu}
\affiliation{Department of Physics, Arizona State University, Tempe, AZ 85287-1504 USA}
\author{Tomoya Takiwaki}
\affiliation{National Astronomical Observatory of Japan, Mitaka, Tokyo 181-8588, Japan}
\author{Tomoya Kinugawa}
\affiliation{Faculty of Engineering, Shinshu University, 4-17-1, Wakasato, Nagano-shi, Nagano, 380-8553, Japan}
\affiliation{Research Center for Advanced Air-mobility Systems, Shinshu University,  4-17-1, Wakasato, Nagano-shi, Nagano, 380-8553, Japan}
\affiliation{Research Center for the Early Universe (RESCEU), School of Science, The University of Tokyo, Bunkyo, Tokyo 113-0033, Japan}
\author{Shunsaku Horiuchi}
\affiliation{Department of Physics, Institute of Science Tokyo, 2-12-1 Ookayama, Meguro, Tokyo 152-8551, Japan}
\affiliation{Center for Neutrino Physics, Virginia Tech, Blacksburg, Virginia 24061, USA}
\affiliation{Kavli IPMU (WPI), UTIAS, The University of Tokyo, Kashiwa, Chiba 277-8583, Japan}
\author{Kei Kotake}
\affiliation{Department of Applied Physics, Fukuoka University, 8-19-1, Nanakuma, Jonan, Fukuoka, 814-0180, Japan}

\date{\today}

\begin{abstract}
We present a new, state-of-the-art computation of the Diffuse Supernova Neutrino Background (DSNB), where we use neutrino spectra from multi-dimensional, multi-second core-collapse supernova simulations --- including both neutron-star and black-hole forming collapses ---  and binary evolution effects from modern population synthesis codes. Large sets of numerical results are processed and connected in a consistent manner, using two key quantities, the mass of the star's Carbon-Oxygen (CO) core at an advanced pre-collapse stage --- which depends on binary evolution effects ---  and the compactness parameter, which is the main descriptor of the post-collapse neutrino emission. 
The method enables us to model the neutrino emission of a very diverse, binary-affected population of stars, which cannot unambiguously be mapped in detail by existing core-collapse simulations. 
We find that including black hole-forming collapses enhances the DSNB by up to $\sim 50\%$ at $E \gtrsim 30$--$40$\,MeV.  Binary evolution effects can change the total rate of collapses, and generate a sub-population of high core mass stars that are stronger neutrino emitters. However, the net effect on the DSNB is moderate -- up to $\sim 15\%$ increase in flux --  
 due to the rarity of these super-massive cores and to the relatively modest dependence of the neutrino emission on the CO core mass. 
The methodology presented here is suitable for extensions and generalizations, and therefore it lays the foundation for modern treatments of the DSNB. 

\end{abstract}


\maketitle

\section{Introduction}

Supernova neutrinos are a missing piece in the rapidly developing field of multimessenger astronomy. While the detection of a neutrino burst from an individual collapsing star remains a once-in-a-lifetime event --- which  occurred only once in recent history, with the historic observation of SN1987A  \cite{Hirata:1987aa,PhysRevLett.58.1494,Alekseev:1987ej}  --- there is confidence that, within a few years, the detection of supernova neutrinos will become commonplace, after the Diffuse Supernova Neutrino Background (\df) \cite{NYAS:NYAS319,Krauss:1984aa} starts to be observed at kiloton detectors~\cite{Super-Kamiokande:2011lwo,Super-Kamiokande:2013ufi,Super-Kamiokande:2021jaq,JUNO:2022lpc,Super-Kamiokande:2023xup,CuestaSoria:2023gss}. 

Besides accelerating experimental progress, the \df\ is of great theoretical significance (see \cite{Beacom:2010kk,Lunardini:2010ab,Mirizzi:2015eza,Ando:2023fcc,Tamborra:2024fcd} for reviews), because it gives a global picture of the \sn\ population in the present universe and at cosmological distances as well. It is also an important probe of particle physics beyond the Standard Model, due to its unique sensitivity to new phenomena (e.g., neutrino decay, or neutrino-dark matter scattering) that develop over cosmological propagation distances (see \cite{Jeong:2018yts,Das:2024ghw,Ivanez-Ballesteros:2022szu,Balantekin:2023jlg,MacDonald:2024vtw} for recent examples).

Predicting the \df\ is challenging due to the myriad of effects that influence it; the main ones being stellar evolution, the physics of collapse simulations, the core-collapse rate history, and neutrino physics. Each element needs to be modeled in detail across the population of progenitor stars, which is currently unfeasible. 

Early computations of the \df\  -- e.g.,  \cite{1986ApJ...302...19W,Malaney:1996ar,HARTMANN1997137,Lunardini:2005jf,Horiuchi:2008jz} -- were simplified, using a single parameterized \sn\ neutrino spectrum, combined with the cosmological star formation history.  Later predictions developed more detail, incorporating the variation of \n\ emission across the population in the form of dependence on the the zero-age main sequence  (ZAMS) mass \citep{Totani:1995rg,Lunardini:2012ne,Horiuchi:2017qja,Priya:2017bmm,Kresse:2020nto,Ekanger:2023qzw} and including black-hole-forming collapses (see, e.g., \citep{Lunardini:2009ya,Mathews:2014qba,Nakazato:2015rya,Horiuchi:2017qja,Moller:2018kpn,Ashida:2022nnv,Ashida:2023heb,Kresse:2020nto,Ekanger:2023qzw,Nakazato:2024gem,Martinez-Mirave:2024zck}). These works used the results of 1D and 2D collapse simulations \citep[e.g,][]{2010A&A...517A..80F,hudepohl_thesis_2013,Nakazato:2013maa,Mirizzi:2015eza,Vartanyan:2023zlb}, which were obtained for certain sets of progenitors where stellar evolution effects were only partially included. Binary evolution effects were recognized as potentially important\footnote{Pioneering works of \citet[][]{Vartanyan:2021dxd,Wang:2024dwq,Schneider:2020vvh} employ some binary effects in the modeling of a supernova neutrino burst. The binary progenitor models are computed in Ref.~\citep{Laplace2021,Schneider:2024mue}. Some of the data is publically avialble  in \url{https://doi.org/10.5281/zenodo.14959596}.} --- since most massive stars are in binary systems \cite{sana12} --- but were not included in \df\ calculations. An exception is the work by Kresse, Ertl and Janka \cite{Kresse:2020nto}, where binary-evolved stars were schematically represented by a set of helium stars, but not based on binary populations \cite{Woosley2019ApJ}. Subsequently, in Horiuchi~{\it et al.}~(2021)~\citep[][hereafter H21]{Horiuchi:2020jnc}, binary effects were taken into account based on binary population synthesis calculations developed by Ref.~\cite{Hurley:2002rf} and updated in Ref.~\cite{Kinugawa:2014zha} (described in more detail below). 

Over time, numerical mappings of the  core collapse parameter space have led to identifying some key parameters - other than the ZAMS mass - that most influence the \n\ emission and connect it with stellar evolution: the mass of the Carbon-Oxygen (CO) core  and the compactness parameter \cite{Limongi:2018qgr,Takahashi:2023jfk}. In particular, there is now consensus that the \n\ emission depends mainly on compactness \cite{OConnor2013ApJ,Nakamura:2014caa,Warren:2019lgb}, and the compactness is roughly determined primarily by the CO core mass 
at the advanced stage \citep[see][for the detail]{Takahashi:2023jfk,Patton:2020tiy}, which is a typical outcome of stellar evolution simulations.  The immediate implication of these advances is the possibility to predict the \n\ emission for a given model of stellar population if its CO core mass distribution is known \cite{Horiuchi:2017qja}. This opens up the possibility to consistently interface two important problems: the modeling of neutrino emission from collapsing stars, and the way supernova progenitors evolve over their lifetime including binary effects. 

\begin{figure*}[htbp]
    \centering
    \includegraphics[width=0.85\linewidth]{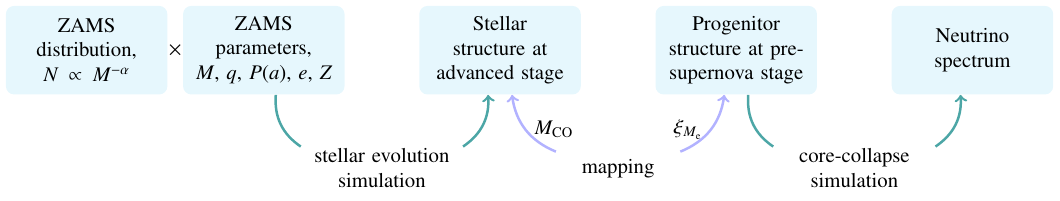}
    \caption{Schematic overview of DSNB flux calculation procedure. The input parameters for stellar evolution simulations are specified at zero-age main sequence (ZAMS) and include the mass $M$, the mass ratio $q$, the orbital period $P$ or separation $a$, the eccentricity $e$ and the metallicity, $Z$. The simulations typically terminate at C or Ne ignition, and provide the CO-core mass, $\mco$. Using stellar profiles given by \emph{other} stellar evolution models (with known $\mco$), core-collapse simulations are performed and provide neutrino spectrum, which is well described by the compactness, $\xi_{M_{\rm e}}$. In our work, we follow \citetalias{Horiuchi:2020jnc} and map CO-core mass and compactness $\xi_{M_{\rm e}}$ to connect the two sets of stellar models, and ultimately obtain the neutrino emission of an evolved population of stars. See text for details.}
    \label{fig:overview}
\end{figure*}

This idea was first applied to the \df\ in the exploratory studies of Horiuchi~{\it et al.}~(2018)~\citep[][]{Horiuchi:2017qja}, where over 100 progenitors from Ref.~\cite{Nakamura:2014caa} was used in determining the DSNB, and in \citetalias{Horiuchi:2020jnc} with a focus on the effect of binary evolution. In \citetalias{Horiuchi:2020jnc}, core-collapse simulations based on 20 progenitors were used from the study of Ref.~\cite{Summa:2015nyk} and binary effects were taken into account based on binary population syntheses calculations. Note that most, though not all, of the core-collapse simulations employed in \citetalias{Horiuchi:2020jnc} typically terminate in a short time, $t <1\,{\rm s}$ \cite{Nakamura:2014caa,Summa:2015nyk}, and their neutrino flux outputs are extrapolated to obtain the whole time evolution of the neutrino emission \cite{Arcones:2006uq}. This introduces a systematic uncertainty \cite{Ekanger:2022neg} which can be updated using long-term simulations \cite{Ekanger:2023qzw}.  
In \citetalias{Horiuchi:2020jnc}, binary effects were modeled using the population synthesis method developed by Ref.~\cite{Hurley:2002rf} and updated in Ref.~\cite{Kinugawa:2014zha} (described in more detail below). The modifications to the CO mass distribution due to binary interactions were quantified using six binary population synthesis realizations, but since the number of supernova progenitors used to study neutrino emission was sparse (20 progenitors), a schematic mapping was developed: the neutrino emission was modeled using the core compactness parameter, and the mapping between core compactness and CO core mass was quantified. This allowed \citetalias{Horiuchi:2020jnc} to model the neutrino emission from hundreds of thousands of supernova progenitors in the binary population synthesis models. See Fig.~\ref{fig:overview} for an illustration of the procedure.

Here we further develop the method of \citetalias{Horiuchi:2020jnc}. We discuss its elements in detail, and to what degree they influence the \df. For each element, we employ the most recent, state-of-the-art predictions. Among these, an important update is the \n\ emission parameters (spectra and luminosities) from the recent set of 100 multi-second 2D core collapse simulations of \citet[][hereafter VB23]{Vartanyan:2023zlb}. We also use the results of two stellar evolution codes, BPASS \cite{Patton:2021gwh} and the code by Kinugawa, Inayoshi, Hotokezaka, Nakauchi,
and Nakamura \cite[][\km\ from here on; adopted in previous study \citetalias{Horiuchi:2020jnc}]{Kinugawa:2014zha}, where binary evolution effects are modeled in detail,
including key processes such as Roche lobe overflow, common envelope evolution, and stellar mergers. These effects can significantly change the final fate of massive stars: for instance, mass transfer can change the stellar masses, and can strip hydrogen envelopes, leading to helium stars or Wolf-Rayet-like progenitors, while stellar mergers can produce more massive cores than single-star evolution would allow. Such processes alter the CO core mass and compactness at collapse, thereby influencing the resulting neutrino emission. The BPASS code and the KIHNN code track these effects across a wide range of initial binary parameters and assumptions of binary interaction parameters.
Our study is meant to lay out a methodology that is particularly well suited for future developments. Indeed, once substantial new sets of numerical simulations are accessible, it will be feasible to derive the fundamental dependencies on CO-core mass and compactness from them, and subsequently update the \df\ calculation in a consistent manner. 

The paper is structured as follows. In Sec.~\ref{sec:overview}, the method is introduced, along with some basic general information.  In Sec.~\ref{sec:NuMcoXi}, the neutrino emission from collapsing stars is described in detail, with emphasis on its dependence on compactness. Sec.~\ref{sec:BinaryEffects} contains a discussion of binary evolution effects. In Sec.~\ref{sec:DSNBresults}, the elements presented in the previous sections are combined into the calculation of the \df, and results are presented. A summary and discussion follows in Sec.~\ref{sec:discussion}.  

\section{Overview and generalities}
\label{sec:overview}

Before delving into a detailed description, here we summarize the main parts of this work and clarify their interplay. We also define the key quantities and notation. 

To calculate the \df, we require: (i) the neutrino flux of individual core-collapses, and  (ii) the distribution of the population of these events. The neutrino flux is given by core-collapse simulations with neutrino radiation transfer and the distribution of the event is provided by stellar evolution calculations. To bridge the two sets of simulations, we have to parametrize the simulation results using compactness and CO-core mass. We illustrate the procedure in Fig.~\ref{fig:overview}. The elements shown in the figure are briefly described below. 

To model the \n\ emission from a supernova, we employ the results of the
most recent, state-of-the-art models: 100
core-collapse simulations developed by \citetalias{Vartanyan:2023zlb}, which are 
two-dimensional~(2D), and extend up to 4\,s post-bounce. 
In these simulations,
black hole formation appears for ZAMS mass  $M\simeq$ 12--15\,$\msun$, and this trend is different from the previously established picture, where black holes were formed for $M\ge 20\,\msun$ \citep[e.g.,][]{Woosley2002RvMP}.
To characterize the neutrino flux,
we use the compactness parameter \cite{OConnor:2010moj}, defined as:
\be
\xi_{M_{\rm e}}=\frac{M_{\rm e}/\msun}{R/1000\,{\rm km}}
\label{eq:xidef}
\ee 
which is evaluated at the pre-collapse phase. This quantity is obtained by integrating the mass density up to a certain radius, $R$, to obtain the mass enclosed within that radius, $M_{\rm e}$.
It is customary to fix the value of $M_{\rm e}$; here we adopt $M_{\rm e}=2.5\msun$, following \citetalias{Horiuchi:2020jnc}. Note that other choices of $M_e$ are correlated with that of $M_{\rm e}=2.5\msun$ \cite{Pejcha:2014wda}, and lead to very similar results in our study.  
The compactness parameter has the merit of being simple, and it is widely used to predict the key observables, such as compact remnant mass, explosion energy and the properties of the neutrino emission \cite{Takahashi:2023jfk,OConnor2013ApJ,Nakamura:2014caa,Warren:2019lgb}.\footnote{However, we note that compactness alone may not be the best predictor of supernova explodability \cite[e.g.,][]{Wang:2022dva,Tsang:2022imn,Burrows:2024pur}.}
We refer to Sec.~\ref{subsec:Xinuparam} for details.

To portray the distributions of the core-collapse events, the mass of the CO core, $\mco$, is a good parameter of choice. The reason is practicality: 
in large stellar evolution grids, evolving all models to the pre-collapse phase is computationally challenging, and obtaining the compactness is difficult. Instead, $\mco$ is more readily available in the most stellar evolution studies \cite[e.g.,][]{Patton:2021gwh,Horiuchi:2020jnc,Fragos2023ApJS}. 
This motivates using a mapping to connect $\xic$ and $\mco$ to connect stellar evolution simulations and core collapse ones (Fig.~\ref{fig:overview}; see Sec.~\ref{subsec:McoXi} for details).
Rigorously, the value of $\mco$ at earlier stages, typically C or Ne ignition --- which is most often provided by stellar evolution simulations --- may slightly differ from that at the onset of core collapse.\footnote{In BPASS, stars are evolved until core-collapse, so this issue does not apply.} Also, the compactness depends on 
the composition of the CO-core \cite{Patton:2020tiy} and this effect is not included in our work. These problems should be addressed in future studies.

A novel aspect of our study lies in the detailed comparison between single and binary star evolution, according to different numerical models.
In Sec.~\ref{sec:BinaryEffects}, we present the distributions of $\mco$ based on two recent stellar evolution models: BPASS~\cite{Patton:2021gwh} and the \km\ model~\cite{Kinugawa:2014zha}, which was employed in \citetalias{Horiuchi:2020jnc}.
These distributions are obtained by assuming certain distribution of initial binary parameters, such as mass ratio, orbital period (separation), and eccentricity.

Considering the cosmological time evolution and effect of neutrino oscillation, we obtain DSNB flux, as described in Sec.~\ref{sec:DSNBresults}.

\begin{figure*}[htbp]
    \centering
    \includegraphics[width=0.85\linewidth]{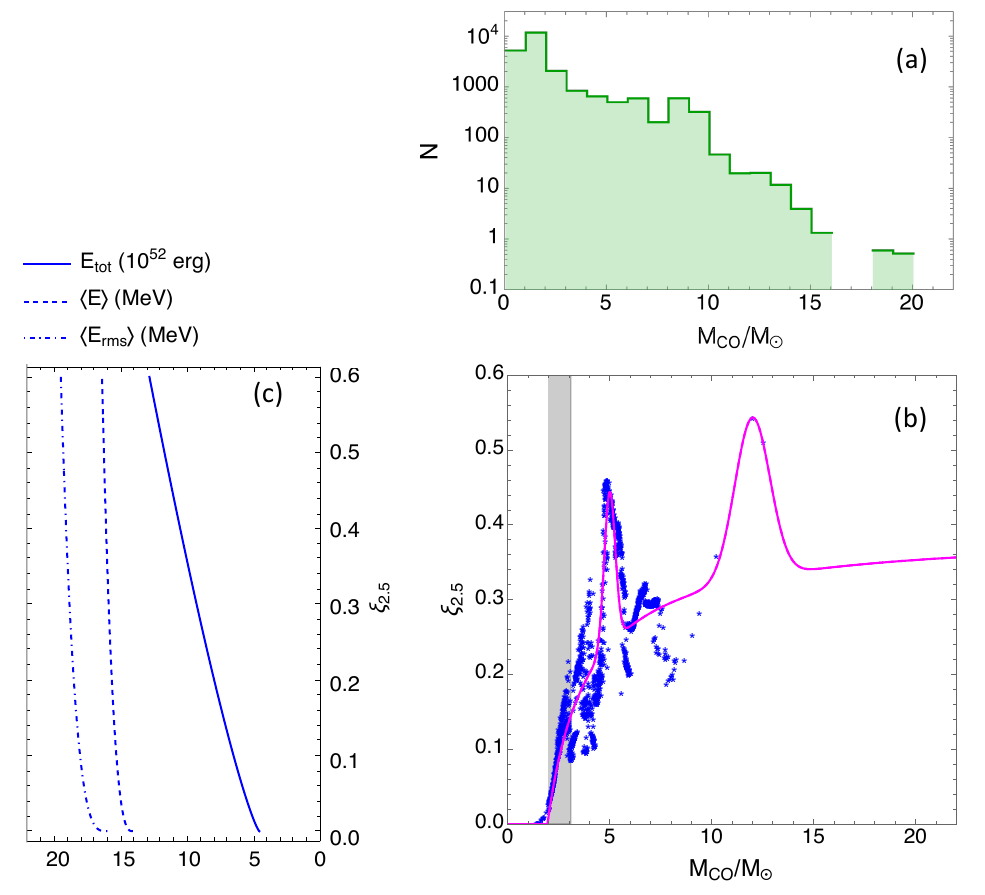}
    \caption{An illustration of the main elements used to compute the population-averaged flux for redshift $z=0$ (and therefore the \df, after accounting the evolution with $z$, see Sec.~\ref{sec:DSNBresults}). (a) The stellar population is described by its distribution  in bins of  $\mco$. Shown here is the distribution obtained using BPASS including binary evolution effects (Sec.~\ref{sec:BinaryEffects}). For each star of given $\mco$, the neutrino emission is modeled considering that: (b) the compactness parameter ($\xic$) is a function of $\mco$ (here numerical results, as well as an interpolating curve, are shown), and (c) the parameters describing the \n\ luminosity and spectrum (for each flavor; here results for $\barnue$ are shown) can be described by simple functions of $\xic$. The shaded area indicates the region ($2.0<\mco/M_\odot <3.1$) where black hole formation was found in the  simulations of \citetalias{Vartanyan:2023zlb}. We note that, in the range $\xic>0.37$ the curves in (c) are an extrapolation; see Fig.~\ref{fig:nuparam} for details. }
    \label{fig:introfigure}
\end{figure*}

\section{Supernova neutrinos: dependence on core masses and compactness}
\label{sec:NuMcoXi}

\subsection{From core mass to compactness}
\label{subsec:McoXi}

We begin by establishing a mapping that relates compactness and CO core mass.
For the dependence of $\xic$ on $\mco$, we adopt a combination of the stellar evolution results of Sukhbold et al. 2016 \cite{Sukhbold:2015wba} and Sukhbold Woosley and Heger 2018 \cite{Sukhbold:2017cnt}, where the most recent results are used in case the two sets overlap (which is for $12  \leq M/\msun \leq 26.99$). In these works, progenitor stars were evolved up to the presupernova stage, defined as the time when the core collapse speed exceeds $v_{\rm coll}=900~{\mathrm{km \, s^{-1}}}$ \cite{Sukhbold:2017cnt}.  The data in the $\xic$--$\mco$ plane are shown in Fig.~\ref{fig:introfigure}. Despite a certain amount of numerical scatter, originated from the non-linear evolution of  C shell burning \citep{Sukhbold:2019kzi}, the main trends are well visible, in particular the increase of $\xic$ in the region $2 \lesssim \mco/M_\odot \lesssim 4 $, and a peak at $\mco \sim 5 \msun$. At higher $\mco$, the data are more sparse, and so a clear pattern can not be identified. Still, there is indication of an increase in compactness with increasing $\mco$, in the region $7 \lesssim \mco/M_\odot \lesssim 12 $. 

For the purpose of computing the \df, we found it convenient to use a functional fit of these numerical data, in the form of a power law  and two gaussian peaks, see Fig.~\ref{fig:introfigure},  and Appendix~\ref{subsub:ximcofun} for details.  The choice of this function is in part informed by others in the literature, in particular \citet{Limongi:2018qgr}, and \citet{Mapelli:2019ipt} for the general increasing trend, and \citet{Sukhbold:2017cnt} for the existence of a second peak at $\mco \sim 12 \msun$. The latter is absent in other works, e.g., \citet{Takahashi:2023jfk}, where a monotonic increase at high $\mco$ appears instead. While we decided to include the second peak in our work, we stress that its presence has a minor effect on the \df, as the region $\mco \gtrsim 10 \msun$ is very sparsely populated, as will be discussed in Sec.~\ref{sec:BinaryEffects}. 
Modeling the high $\mco$ region as a narrow peak leads to a slightly lower diffuse \n\ flux (because the neutrino flux tends to increase with $\xic$, see Sec.~\ref{subsec:Xinuparam} below) compared to a model with a monotonic increase that is extrapolated to $\mco\gtrsim 12 \msun$.

\begin{figure}
    \centering
    \includegraphics[width=0.95\linewidth]{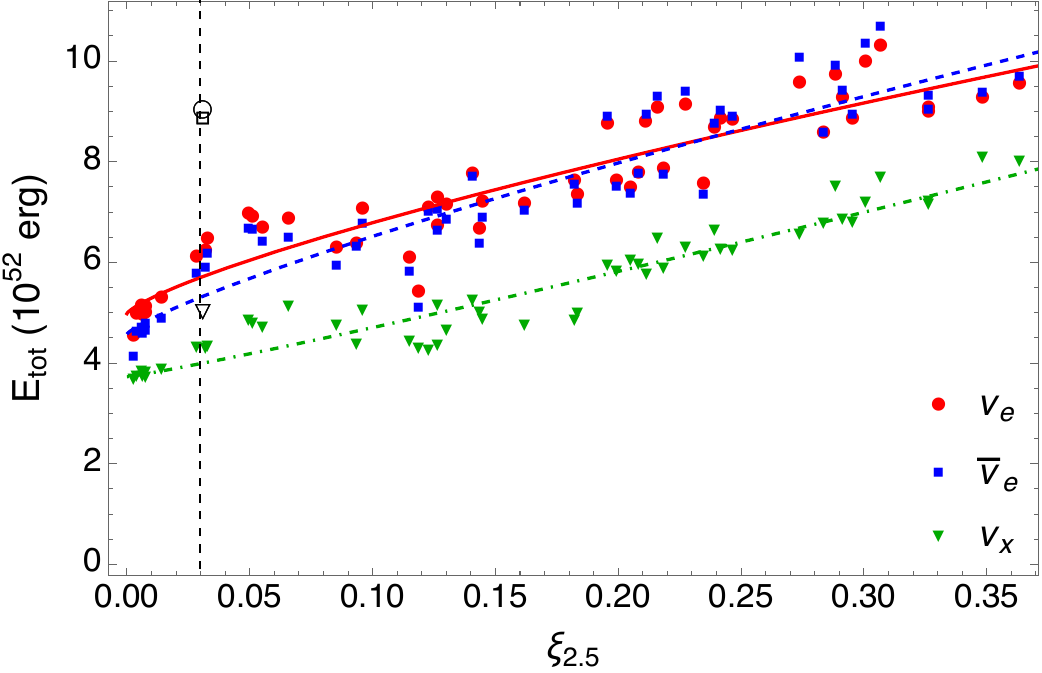}
    \includegraphics[width=0.95\linewidth]{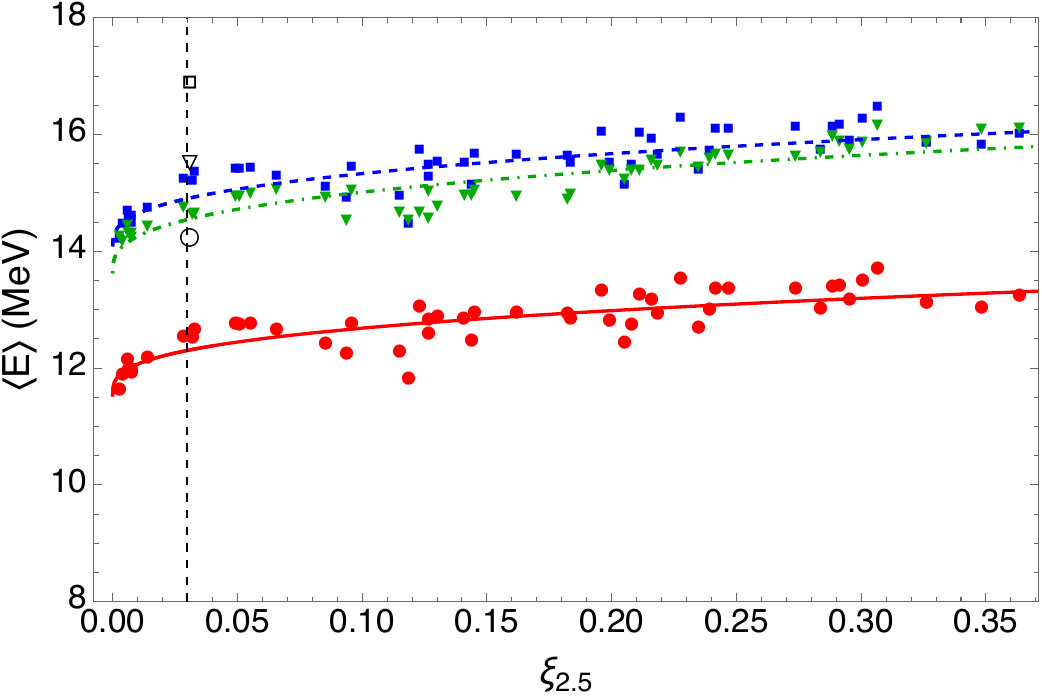}
    \includegraphics[width=0.95\linewidth]{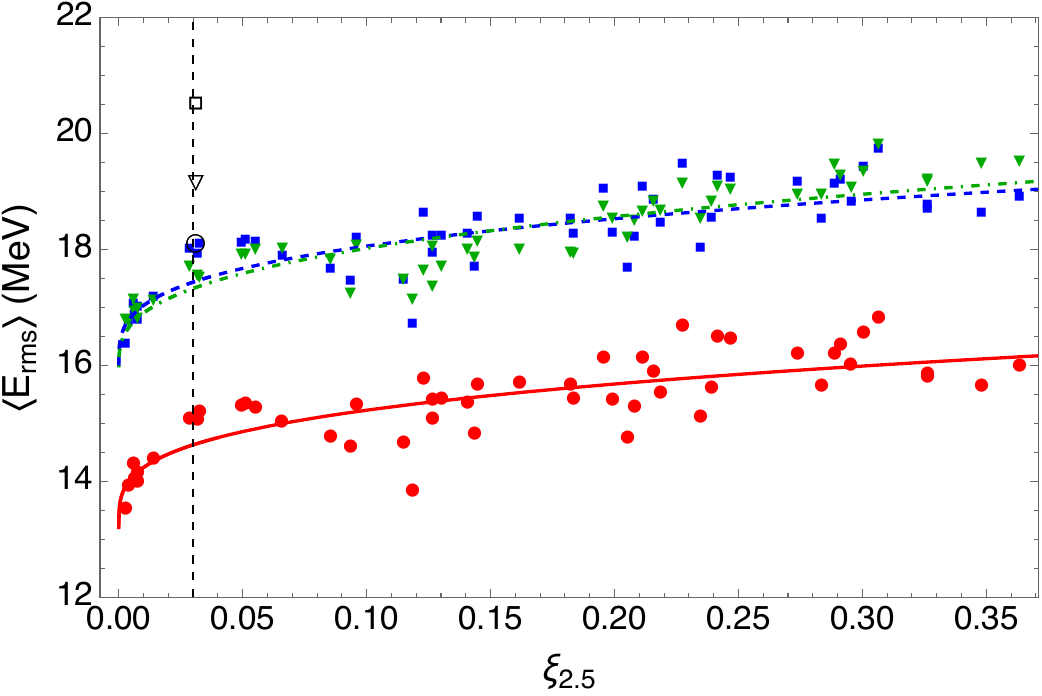}
    \caption{Neutrino parameters as functions of $\xic$, from the results of \citetalias{Vartanyan:2023zlb}, for runs that extend to $t\geq 3.5$\,s post-bounce.  For each flavor we show the total energy emitted (over the entire burst, obtained by extrapolation), the average energy, and the rms energy. The filled markers in color are for neutron-star-forming collapses; for these, functional fits are shown (curves of matching color). The empty markers along the vertical dashed line refer to black-hole-forming collapses.  }
    \label{fig:nuparam}
\end{figure}

\subsection{From compactness to neutrino flux parameters}
\label{subsec:Xinuparam}

We now examine the results of the core collapse simulations in \citetalias{Vartanyan:2023zlb}, 
to describe the dependence of the \n\ flavor fluxes on $\xic$. 
These simulations employ  stellar progenitors models from \cite{Sukhbold:2015wba,Sukhbold:2017cnt}, and therefore they are consistent with our choice for the dependence of $\xic$ on $\mco$ (Sec.~\ref{subsec:McoXi}). 
For most progenitors, a successful shock revival is obtained by the neutrino-heating mechanism, leading to exploding, neutron-star forming collapses (\nsfc). For most models in the region $12 \leq M/M_\odot \leq 15.38$ (corresponding to $2.0<\mco/M_\odot <3.1$),  however,  the neutrino-driven shock revival was not obtained, leading to black hole formation (black-hole-forming collapses, \bhfc). 

We used the publicly available tables in \citetalias{Vartanyan:2023zlb}, where detailed information is given for each neutrino species: electron neutrinos ($\nue$), electron antineutrinos ($\barnue$) and the muon and tau neutrinos and antineutrinos (these being represented as a single effective species, $\nux$). Specifically, for each (effective) \n\ species and each instant of time, the table gives the instantaneous luminosity, and the first three moments of the energy spectrum --- in other words, the average energy, $\langle {\mathcal E}\rangle$, the root mean square energy, $\langle {\mathcal E}_{\rm rms} \rangle \equiv (\langle {\mathcal E}^2 \rangle)^{1/2}$, and $(\langle {\mathcal E}^3 \rangle)^{1/3}$ --- . 
To ensure internal consistency, out of the 100 simulations results, we restricted to those that extended to at least 3.5\,s post-bounce. This subset includes 53 progenitors, of which 52 evolve into \nsfc\  and 1 (with $M=12.1\msun$ and $\mco=2.09\msun$) yields a \bhfc. These models are characterized by long-lived proto-neutron stars where mass accretion continues for several seconds before collapse to a black hole. The late-time tail of the neutrino emission is relevant for their contribution to the total flux. 
Unlike the exploding case, where analytic prescriptions for the late-time neutrino luminosity are available~\cite{Suwa:2020nee}, no analogous description exists for BHFC, and current approaches only treat quasi-steady accretion states~\cite{Akaho:2023alv}.
Using the longest available BHFC model minimizes the uncertainties associated with truncating the accretion-powered neutrino emission.

By performing the appropriate integrations, we obtained the quantities describing the time-integrated (over the duration of the simulation) \n\ emission, specifically the total energy emitted in a given flavor  ($w=e,\bar e,x$, referring to $\nue,\barnue, \nux$) $E_{{\rm tot},w}$, and the average and rms energies: $\langle E\rangle_w$ and $\langle E_{\rm rms}\rangle_w$. As a second step, we applied a correction to $E_{{\rm tot},w}$ to include the energy emitted at late times, beyond the end of the simulation. The correction was done by estimating the \emph{total} (of all times) energy emitted by fitting the luminosity as a function of time with a negative exponential function. We find that these energies corrected by extrapolation are higher by 0 -- 35\%  than the ones obtained by integrating over 3.5\,s. A $\sim 20\%$ correction is the most common outcome, and larger (smaller) corrections are generally observed for $\nux$ ($\nue$) ~\footnote{We decided not to apply a similar correction to $\langle E\rangle_w$ and $\langle E_{\rm rms}\rangle_w$ because, in several cases, the instantaneous average and rms energies remain constant in time or even increase over the duration of the simulation. The lack of a clear trend made it unadvisable to extrapolate.}.

Results for the neutrino parameters are shown in Fig.~\ref{fig:nuparam}.  In the figure, some general trends are visible that are consistent with previous literature; in particular we notice that the spectral parameters, as well as the total emitted energy, increase with compactness. The average and rms energies show the well established hierarchy between $\nue$ and $\nux$, with $\nue$ having softer spectrum.  
Instead, in the antineutrino sector there is no strong hierarchy: the rms energies are very similar, and the $\barnue$ species has slightly (by $\sim 0.5$ MeV or so) higher everage energy than $\nux$, which is in contrast with the traditional trend ($\nux$ having hotter spectrum) found in most literature.
The total energy emitted in neutrinos is not equipartitioned between the flavors: the $\nux$ flux (each of the 4 species that are labeled $\nux$) carries less energy away from the proto-neutron star. For black-hole-forming collapses, the \n\ emission is more energetic: for each flavor all energy parameters are higher than for neutron-star forming collapses of similar compactness.

In Fig.~\ref{fig:nuparam}, the lower luminosity of $\nu_x$ compared to the electron flavors is a robust feature across other simulations. However, the \citetalias{Vartanyan:2023zlb} dataset shows the unusual trend that the average energy of $\nu_x$ is slightly smaller than that of $\bar{\nu}_e$, while most other studies find the opposite ordering \cite[e.g.,][]{Nakazato:2013maa,Horiuchi:2017qja}. This behavior may be related to the longer simulation time of \citetalias{Vartanyan:2023zlb}  ($\sim 4\,{\rm s}$), during which the spectra of different flavors tend to converge, or it could be an artifact of the approximate neutrino transport scheme employed. More sophisticated simulations with improved transport and longer durations will be needed to clarify this point.

For \nsfc, the  dependence of  $E_{{\rm tot},w}$, $\langle E\rangle_w$ and $\langle E_{\rm rms}\rangle_w$  on $\xic$  was found --- by numerical fitting ---  to be well described by a function of the form $a + b(\xic)^c$, which is shown in Appendix~\ref{subsub:nuxifun}.    
 The dependence is approximately linear ($c\sim 1$) for $E_{{\rm tot},w}$, and like the cubic root ($c\simeq 1/3$) for $\langle E\rangle_w$ and $\langle E_{\rm rms}\rangle_w$.~\footnote{In Fig.~\ref{fig:nuparam}, it is possible to see a dip at $\xic\simeq 0.12$ in the numerical neutrino parameters that is not well reproduced by the fitting functions. We found that this dip corresponds to a region in the $\xic$--$\mco$ plane where there is a significant numerical scatter (see Fig.~\ref{fig:introfigure}). Therefore, we chose to consider it as non-significant. }  One may wonder if there is a  physical reason behind these trends; we were unable to find one, and we leave this intriguing question for future investigation. The fitted curve is almost consistent with Fig.~5 of \citetalias{Horiuchi:2020jnc} though the employed parameters are slightly different. 
 
 For the non-exploding case, the parameters of the single simulation we used (see Fig.~\ref{fig:nuparam}), are taken to be representative of the whole population of \bhfc.~\footnote{We verified that this is reasonable, because 
neutrino parameters vary minimally over the black-hole forming subset (for early times, when several BH-forming simulation results are available), mostly within a few per-cents. A maximum variation of $\sim 20\%$ was seen, for the total energy emitted in $\nue$. } 
This treatment significantly differs from the previous works. For instance, \cite{Horiuchi:2017qja} assumes that black hole formation occurs above a critical compactness with the neutrino emission modeled in the range of $0.2 \le \xi_{2.5} \le 0.8$ based on multiple 1D simulation models (see their Fig.~5). However, in multi-dimensional simulations, the explodability of supernovae is not determined by such a simple criterion and progenitors within this range can still explode, see Ref.~\cite{Vartanyan:2023zlb,Choi:2025igp} for example. A compilation of black hole formation models can be found in Ref.~\cite{Suwa:2025oad}.

From $E_{{\rm tot},w}$, $\langle E\rangle_w$ and $\langle E_{\rm rms}\rangle_w$, the (time-integrated) \n\ spectrum for the species $w$ (i.e., the number of \n\ of species $w$ emitted per unit energy) were modeled using the well known alpha spectrum \cite{Keil:2002in,Tamborra:2012ac}:  
\begin{align}
    F_w(E,\xic)=  \frac{(1+\alpha_w)^{1+\alpha_w}E_{{\rm tot},w}}
  {\Gamma (1+\alpha_w){\langle E\rangle_w}^2}
  \left(\frac{E}{{\langle E\rangle_w}}\right)^{\alpha_w}\nonumber\\
  \times\exp\left[{-(1+\alpha_w)E/{\langle E\rangle_w}}\right],
  \label{eq:alphaspectrum}
\end{align}
where $E$ is the neutrino energy, 
 and $\alpha_w$ is a numerical parameter describing the shape of the spectrum; it is defined by the equation $(2+\alpha_w)/(1+\alpha_w)\equiv\langle E_{\rm rms}\rangle^2_w/\langle E\rangle^2_w$. Here $F_w$ depends on $\xic$ through the functions given in Fig.~\ref{fig:nuparam}; we find that $\alpha_w$ (not shown in Fig. \ref{fig:nuparam}) decreases with increasing $\xic$, and varies between 1.0 and 2.1. 

We note that instantaneous \n\ spectra are included in the public data set of \citetalias{Vartanyan:2023zlb}, therefore, one could use them to obtain the time-integrated flavor spectra. We have tested these numerically computed spectra,  and we found that, in most cases, they are in good agreement with those modeled using Eq.~\eqref{eq:alphaspectrum}\cite[see also][]{Tamborra:2012ac}. However, there are several instances where the numerical spectrum is significantly higher than the alpha spectrum at $E \gtrsim 50$\,MeV or so. The difference is a factor of several at 50--80\,MeV, and it can reach several orders of magnitudes at $E \gtrsim 100$\,MeV. This high energy tail is more pronounced for $\nue$, but is observed for all the flavors.  In our understanding, its physical origin, as well as the size of the numerical uncertainties that contribute to it, are not completely clear at this time \cite{VandBprivcomm}. Therefore, we opted for using the alpha spectrum throughout this work, thus potentially underestimating the \df\ at the highest energies.

\begin{figure*}[htbp]
    \centering
    \includegraphics[width=0.80\linewidth]{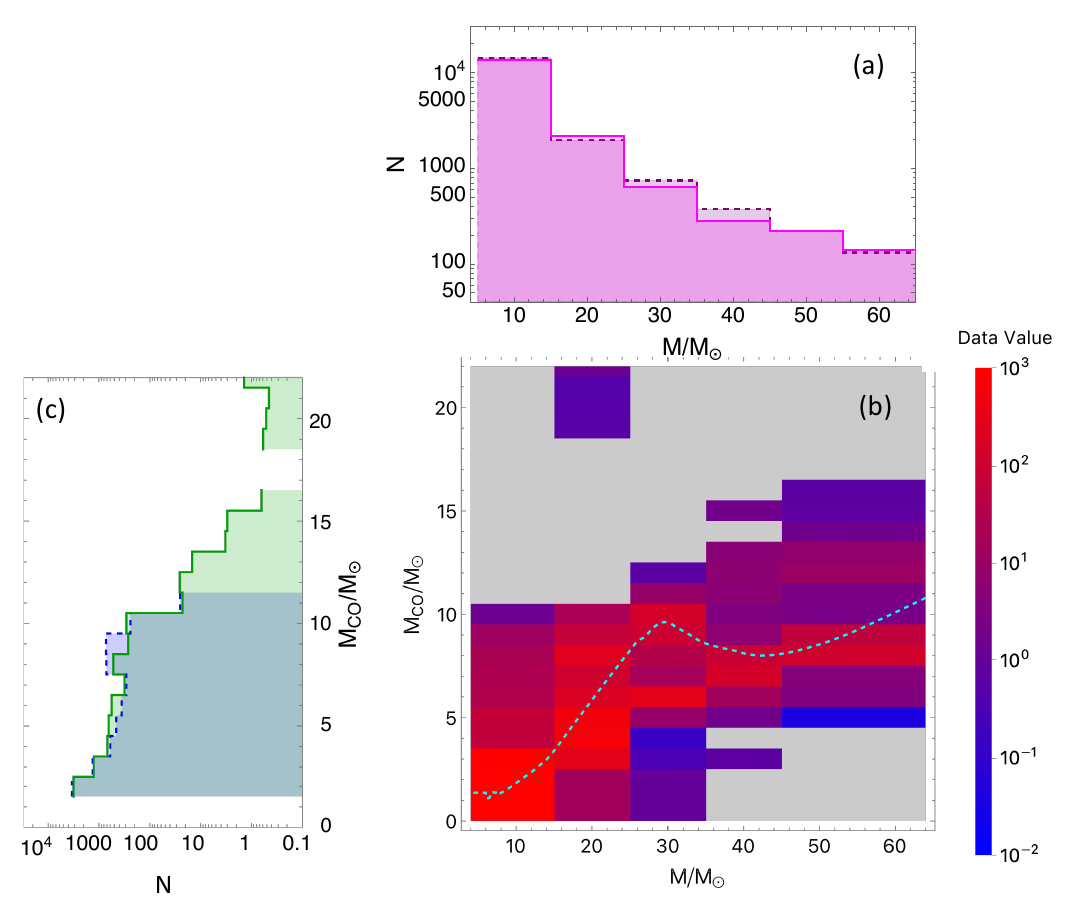}
    \caption{  Overview of BPASS results, for single and binary stars (dashed and solid lines respectively).  In (a) and (c), $N$ indicates the number of stars in a bin of either ZAMS mass ($M$) or CO core mass ($\mco$). Panel (b) shows a density plot, where
each pixel represents a bin in the $\mco-M$ parameter space, for binary stars. The colors represent the number of stars in each bin, see legend  (numbers are not integer due to some integration or averaging procedures involved in BPASS \cite{PattonPrivComm}). The gray color indicates $N=0$ (empty bin).  The cyan dashed line shows the dependence of $\mco$ on $M$ for single stars. The total number of stars is $N_{\ast}=24038.5$ and $N_{\ast}=22745.6$ for single and binary stars respectively. Note that we  restrict to the interval $4.1 
\leq M/\msun \leq 65$ to ensure consistency between the single and binary sets.  }
    \label{fig:BPASSoverview}
\end{figure*}

\begin{figure*}[htbp]
    \centering
    \includegraphics[width=0.80\linewidth]{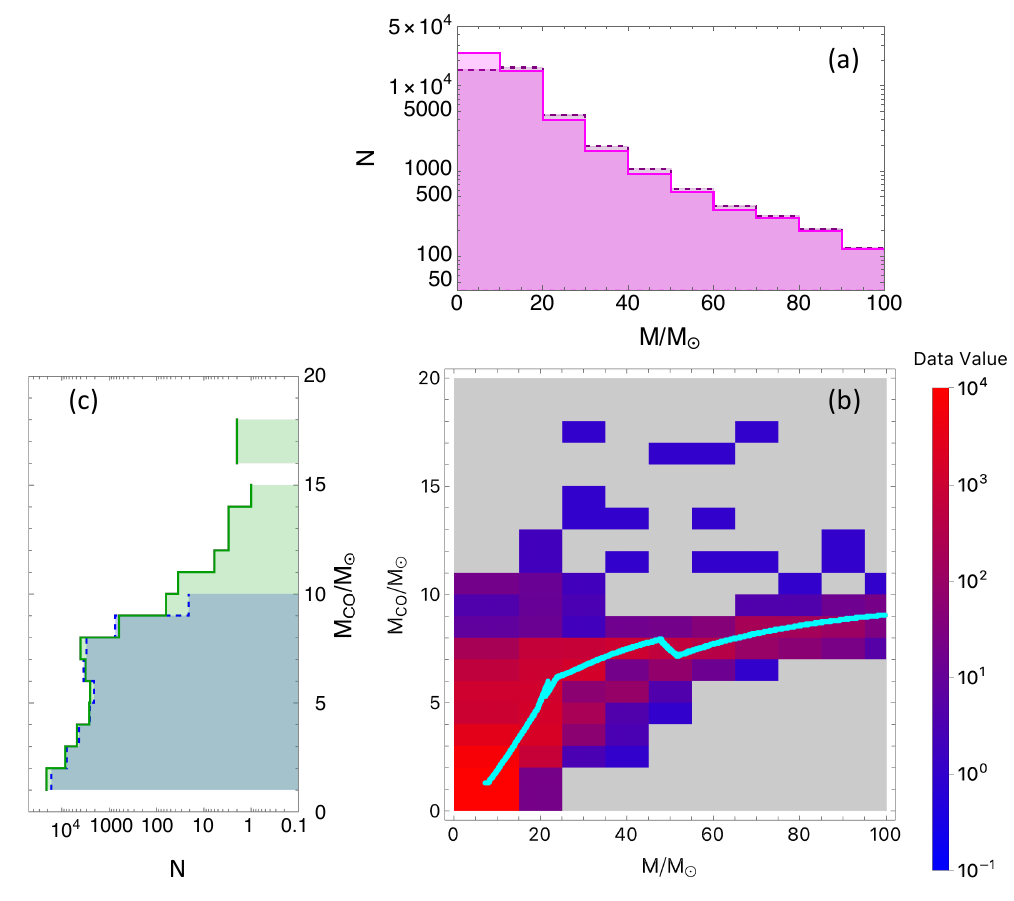}
    \caption{The same as Fig.~\ref{fig:BPASSoverview}, for the fiducial \citetalias{Kinugawa:2014zha} model. The total number of stars in this data set is $N_{\ast}=41066$ and $N_{\ast}=47529$ for single and binary stars respectively.} 
    \label{fig:BPASSoverviewKin}
\end{figure*}

\section{Population synthesis: binary evolution effects}
\label{sec:BinaryEffects}
Binary star evolution can involve a wide range of physical processes that are absent in single star evolution, and these processes can significantly alter the outcome of stellar evolution. One of the most common is mass transfer, in which material flows from one star to its companion, often through Roche lobe overflow. Another key process is common envelope evolution, where both stars share a single, extended envelope and spiral toward each other, potentially leading to a merger. Stellar mergers can produce unusually massive stars with higher core masses than possible via isolated evolution. Binary interactions can also affect angular momentum, rotation, and core structure, which in turn influence the final compact remnant and its neutrino emission.

Because binary evolution proceeds differently depending on the initial parameters of the system, such as the primary mass, mass ratio, orbital separation, and eccentricity, it is necessary to investigate these outcomes statistically by sampling from distributions of these initial parameters. This approach is known as binary population synthesis, and it enables a systematic study of the diverse evolutionary paths and their effects on observable quantities such as the diffuse supernova neutrino background. In addition, binary interactions are influenced by physical processes whose outcomes remain uncertain, such as the accretion efficiency of mass transfer ($\beta$) and the efficiency of common envelope ejection ($\alpha\lambda$), making parameterized treatments necessary within population synthesis calculations.

We adopt two binary population synthesis results.
One is the BPASS v2.2 model~\cite{Patton:2021gwh} and the other is the \km\  model~\cite{Kinugawa:2014zha}. In the BPASS runs, a total mass $M_{\rm tot}=10^{6} \msun$ of single stars, and an equal mass of binary systems are considered.
BPASS is a hybrid Binary Population Synthesis (BPS) code. It evolves its single stars and the primaries of binary stars with a detailed stellar evolution code, while approximating the evolution of the secondaries using the semi- analytic evolution equations from \citet{Hurley2002}. It then evolves the secondaries in detail once the primary star’s evolution terminates.

A detailed description of BPASS is given in \citet{Stanway2018MNRAS}. In the single star models, the ZAMS mass range span $0.2$--$100\msun$ in steps of $0.1\msun$; in the binary models masses span $4.5$--$100\msun$ 
for the primary star's mass, $M_1$, with increments of $0.5\msun$. For the binary systems, the mass ratio, 
$q$, varies between 0.1 and 0.9 in increments of 0.1, and the orbital period, $P$, ranges between $10^{1.4}$ and $10^4$\,days in steps of 0.2 dex in log space.  The distributions of $q$ and $P$ follows \citet{Moe2017ApJS}. All models are for solar metallicity. Stars are evolved until core carbon ignition. BPASS assumes a \citet{Kroupa:2000iv} initial mass function (IMF) for masses between 0.5 and $1\msun$, and a \citet{Salpeter:1955it} IMF  (which follows $M^{-2.35}$) \citep{Salpeter:1955it} for stars with masses heavier than $1\msun$. The mass and radius of the CO-core are computed by fixing the boundary of the CO-core at the radius where the helium mass fraction equals 0.1.

The \citetalias{Kinugawa:2014zha} model adopts the fiducial model ($\alpha\lambda=1$) from \citetalias{Horiuchi:2020jnc} and uses the Abt\_MT1\_CE1 model as described in \citet{Kinugawa2024}.\footnote{The fiducial model ($\alpha\lambda=1$) in \citetalias{Horiuchi:2020jnc} assumes a primary star mass range of 3~$M_{\odot}$ to 140\,$M_{\odot}$, while the Abt\_MT1\_CE1 model in \citet{Kinugawa2024} assumes 3\,$M_{\odot}$ to 100~$M_{\odot}$. However, since the IMF is Salpeter, the contribution from stars in the 100\,$M_{\odot}$ to 140\,$M_{\odot}$ range is negligible, making the two models effectively equivalent. In this study, we adopt the range of 3~$M_{\odot}$ to 100\,$M_{\odot}$.}
The \citetalias{Kinugawa:2014zha} model is calculated by a population synthesis code based on an updated 
version of the binary stars evolution code by \citet{Hurley2002}, 
with details described in Ref.~\cite{Kinugawa:2014zha,Horiuchi:2020jnc,Kinugawa2024}. The main differences include the treatment of mass transfer stability, the handling of the common envelope phase, and the evolution of the cores of rotating stars.

In the \citetalias{Kinugawa:2014zha} model, we assume solar metallicity and adopt a Salpeter IMF for the mass of primary stars in the range of $3$ to $100\,M_\odot$, a constant initial mass ratio function (IMRF) \citep{Kobulnicky:2006bk}, a logarithmically uniform initial separation function (ISF) proportional to the inverse of the separation, $1/a$, \citep{Abt1983}, and an initial eccentricity function (IEF) proportional to the eccentricity, $e$ \citep{Heggie:1975tg}. In addition, pulsar kick velocities are modeled using a Maxwellian distribution with a dispersion of $\sigma = 250\,\mathrm{km\ s}^{-1}$ \citep{Hobbs:2005yx}.
In the \citetalias{Kinugawa:2014zha} model, we use binary parameters set as $(\beta, \alpha \lambda) = (1, 1)$. Here, $\beta = 1$ signifies that no mass is lost from the binary system during the mass transfer. The quantity $\alpha \lambda$ are the parameters of  Common Envelope (CE) phase, where $\alpha$ is the efficiency with which orbital energy is used to unbind the envelope, and $\lambda$ characterizes the binding energy of the envelope. In the \citetalias{Kinugawa:2014zha} model, $\alpha\lambda$ is treated as a single parameter. This parameterization plays a crucial role in determining whether the binary survives the common envelope phase or merges prematurely; see \cite{Horiuchi:2020jnc} for details. When the accretor is a compact object, accretion is limited by the Eddington limit, and any excess mass is ejected from the system. 
If a stellar merger occurs due to a common envelope phase before the star becomes a compact object, the resulting star is treated as a highly rotating star. In this case, the increase in core mass due to rotation is taken into account \cite{Horiuchi:2020jnc,Kinugawa2024}.
A total of $10^5$ binary systems were simulated using this configuration. Note that, differently from BPASS, in the \citetalias{Kinugawa:2014zha} model the sample of stars is not normalized by its mass. 

Fig.~\ref{fig:BPASSoverview} and \ref{fig:BPASSoverviewKin} show the CO core distributions obtained by the two models, for single and binary stellar evolution. 
Despite the differences that exist between different simulations, some predictions are robust. One of them is that, as a result of  binary case, stars are produced with high $\rm CO$ core mass, $\mco \gtrsim 10\msun$, that could not be realized for single star evolution.\footnote{Note that this feature depends on the treatment of mass loss, see Figure~4 of \citet{Fragos2023ApJS} and Figure~1 of \citet{Patton:2021gwh}} These fall in the region of medium-high compactness (see Fig.~\ref{fig:introfigure}~(a)), and therefore have a larger \n\ emission (high $E_{\rm tot}$) than most neutron-star-forming single-evolved stars. In particular, binary-evolved stars populate the (possible) second peak of the $\xic$ distribution ($\mco\sim 12\msun$, see Fig.~\ref{fig:introfigure}~(b)), where $\barnue$s could be emitted with a factor of $\sim 2$ larger total energy ($E_{\rm tot,\barnue}\simeq 1.2 \times 10^{53}\,{\rm erg}$) and a $\sim10\%$ higher average energy than for a \nsfc\  with $\mco\simeq 2.5\msun$.  These progenitors with $\mco \gtrsim 10\msun$ 
 are exceedingly rare, however, amounting to $\sim 1\% $ or less of the entire population.
 
 Another prediction is a wide distribution in $\mco$ that becomes possible from binary evolution even for progenitors of relatively modest ZAMS mass. For example, a binary system where $M_1<M_2\simeq 15\msun$ -- for which single evolution predicts $\mco \lesssim 4 \msun$--  could lead to the formation of a CO core with up to $\mco \simeq 10\msun$. Interestingly, however, the distributions of the stellar population with $\mco$ are relatively similar for the two cases of single and binary evolutions. For the BPASS results, the main difference is that for binary evolution the local maximum at $\mco\sim 8$--$10\msun$ becomes de-populated, whereas the distribution remains practically the same as the single evolution case in the region $\mco \lesssim 4 \msun$. In the \citetalias{Kinugawa:2014zha} model, instead, a slight overpopulation of the low $\mco$ region is found for binary evolution.

\section{Putting it all together: diffuse neutrino flux}
\label{sec:DSNBresults}

We now combine the results of the previous sections with the cosmological rate of core collapse, to compute the \df. 

For a given population synthesis model (BPASS or \citetalias{Kinugawa:2014zha} model), the first step 
is to compute the following fraction:
\begin{equation}
f^{w}_i(E)=\frac{\int^{M_{\rm CO,max}}_{M_{\rm CO,min}} F_w(E,\mco) \left( \frac{dN}{d\mco}\right)_{i} dM_{\rm CO}}{\int^{M_{\rm CO,max}}_{M_{\rm CO,min}}  \left( \frac{dN}{d\mco}\right)_{S} dM_{\rm CO}}~,
    \label{eq:popavg}
\end{equation}
where the index $i$ indicates the case of single or binary star evolution ($i=S,B$) and $F_w(E,\mco)$ is the individual star \n\ emission in the species $w$, which is derived by combining Eq.~\eqref{eq:alphaspectrum} with the function $\xic=\xic\!(\mco)$ as shown in Fig.~\ref{fig:introfigure} (b).
The quantity $\left( \frac{dN}{d\mco}\right)_{i}$ is the distribution of stars in $\mco$, which is displayed in Fig.~\ref{fig:BPASSoverview}~(c), and Fig.~\ref{fig:BPASSoverviewKin}~(c) for BPASS and \citetalias{Kinugawa:2014zha} model, respectively. 

For $i=S$, the expression in Eq.~\eqref{eq:popavg} is the population-averaged neutrino flux, whereas for $i=B$, it is the average multiplied by a normalization factor that accounts for binary evolution effects changing the total number of stars that undergo core collapse. We study two cases: the first is one where all the collapses form neutron stars, therefore $\xic=\xic\!(\mco)$ as shown in the curves in Fig.~\ref{fig:nuparam}. The second case is with black hole formation in the range $2.0 <\mco/M_\odot < 3.1$ (see Sec.~\ref{subsec:Xinuparam}), which corresponds to $\sim 20\%$ of all collapses forming black holes. Here, the neutrino spectrum parameters are taken to be fixed to their \bhfc\ values (empty markers in Fig.~\ref{fig:nuparam}) in the interval $2.0 <\mco/M_\odot < 3.1$, and to the \nsfc\ curves in the complementary region of $\mco$.  

Following \citetalias{Horiuchi:2020jnc}, we make the assumption that $\left( \frac{dN}{d\mco}\right)_{i}$ --- and therefore $f^{w}_i(E)$ --- does not depend on the redshift $z$. This is justified by the fact that  binary evolution effects occur on a time scale which is much shorter than the lifetime of a supernova progenitor, even when binary effects are included. 

For the cosmological rate of core collapses, we use the parametrization in \cite{Yuksel:2008cu} (which was also used in \citetalias{Horiuchi:2020jnc}):
\begin{equation}
\dot{\rho}_{\rm CC}(z) = \dot{\rho}_0 \left[ (1 + z)^{a \eta} + \left( \frac{1 + z}{B} \right)^{b \eta} + \left( \frac{1 + z}{C} \right)^{c \eta} \right]^{1/\eta}~,
    \label{eq:snr}
\end{equation}
  where  $\dot{\rho}_0 = 1.3 \times 10^{-4} \,  \text{yr}^{-1} \, \text{Mpc}^{-3}$, $a = 3.4$, $b = -0.3$, $c = -3.5$, $B=5000$, $C=9$ and $\eta = -10$. 
We checked that using other functional forms (e.g., the ones in \cite{Ekanger:2023qzw}) leads to nearly identical results if the normalization $\dot{\rho}_0$ is the same.

From the quantities above, one obtains the \df\ for a single \n\ species \emph{without} flavor oscillations: 
\begin{equation}
\Phi_0^{w,i}(E) = \frac{c}{H_0} \int_0^{z_{\text{max}}} \frac{ \dot{\rho}_{\rm CC}(z) f_i^w(E') }{\sqrt{\Omega_{\rm m}(1 + z)^3 + \Omega_\Lambda}}dz~,
\label{eq:dsnb}
\end{equation}
where $E' = E(1 + z)$; the fractions of cosmological energy density in matter and dark energy are  $\Omega_{\rm m}=0.3$ and $\Omega_\Lambda=0.7$, and the Hubble constant is fixed to $H_0=71\,{\rm km}\,{\rm s}^{-1} {\rm Mpc}^{-1}$. We fixed $z_{\rm max}=4.5$; due to the fast decline of the supernova rate at $z\gtrsim 4$, the result of the integral depends only minimally on the value of this parameter. Here $c$ in the equation is the speed of light.

To compute the realistic flux of $\barnue$ and $\nue$ in a detector on Earth, one must include the effect of \n\ flavor oscillation. This is done by introducing an effective $\barnue$ survival probability, ${\bar p}$, and writing the (oscillated) diffuse $\barnue$ flux as:
\begin{equation}
\Phi^{\bar e,i}(E) = {\bar p} \Phi_0^{\bar e,i}(E) + (1-\bar p) \Phi_0^{x,i}(E)~.
    \label{eq:dsnboscill}
\end{equation}
A similar expression as Eq.~\eqref{eq:dsnboscill} holds for $\nue$, with the replacement $\bar p \rightarrow p$. The parameters $\bar p$ and $p$ can be computed by modeling the matter-driven and neutrino-driven (i.e., collective flavor oscillations) inside the star, as well as oscillations inside the Earth. Their values are uncertain, due to the uncertainty in the treatment of collective flavor oscillations. For our purpose, it is sufficient to approximate them as constants, with values falling in intervals $\bar p = 0$--$0.7$ and $p=0$--$0.3$. The extremes of these intervals are (approximately) the predictions of matter-driven conversion depending on the choice of neutrino mass ordering (specifically, $\bar p \simeq \sin^2 \theta_{13}$--$\cos^2 \theta_{12}$  and $p\simeq \sin^2 \theta_{13}$--$\sin^2 \theta_{12}$ see, e.g., \cite{Lunardini:2012ne} for a discussion and further references).

\begin{figure*}
  \centering
 {\normalsize\textbf{BPASS  \hskip 5truecm \km\ model}}\\[0.5em]
  \includegraphics[width=0.4\linewidth]{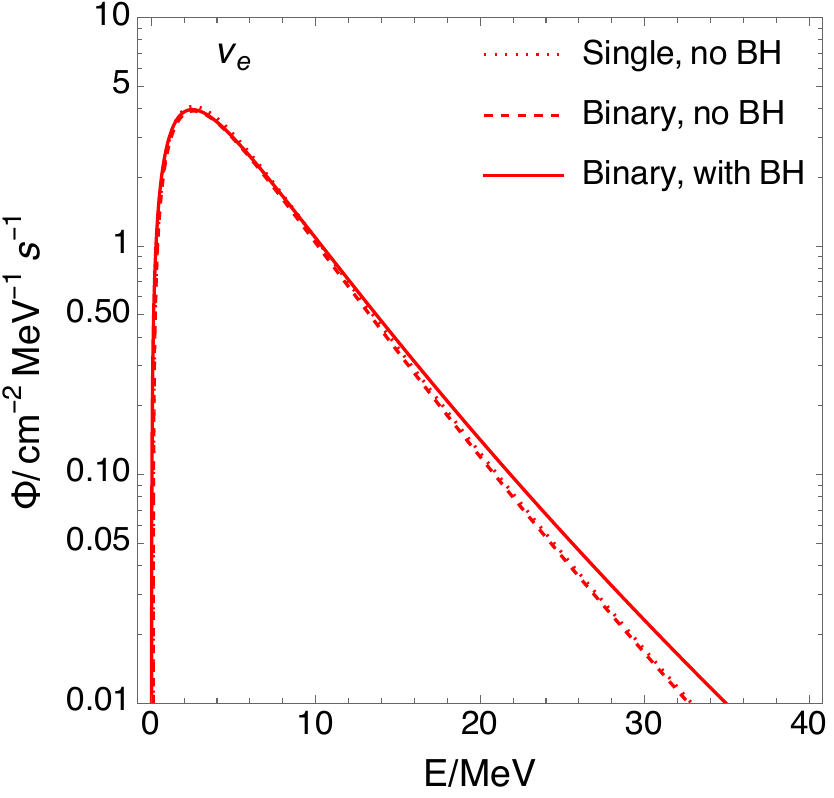} 
  \includegraphics[width=0.4\linewidth]{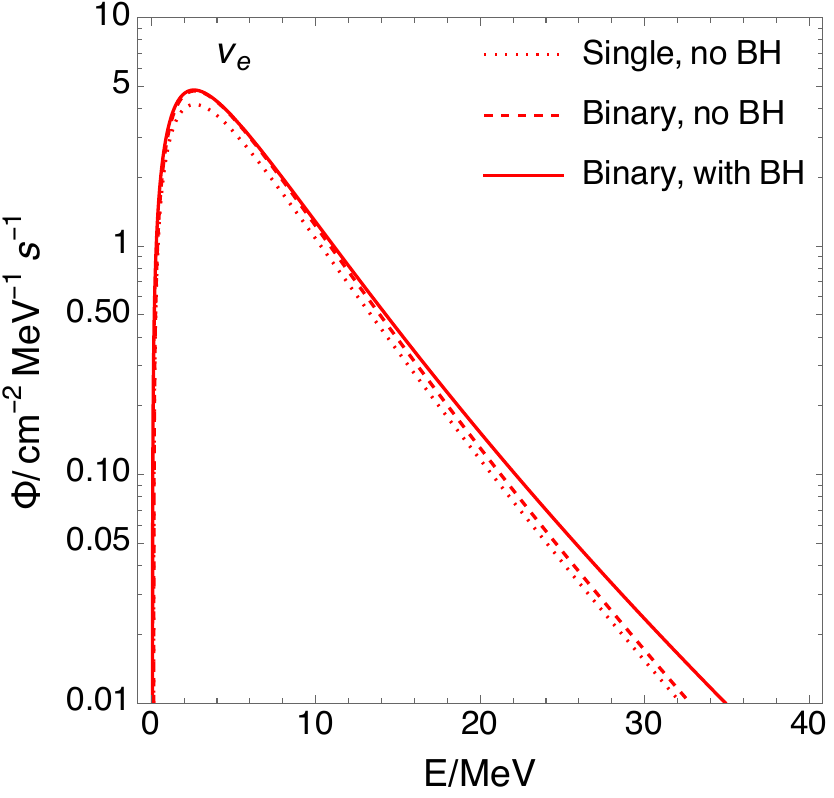}
  \includegraphics[width=0.4\linewidth]{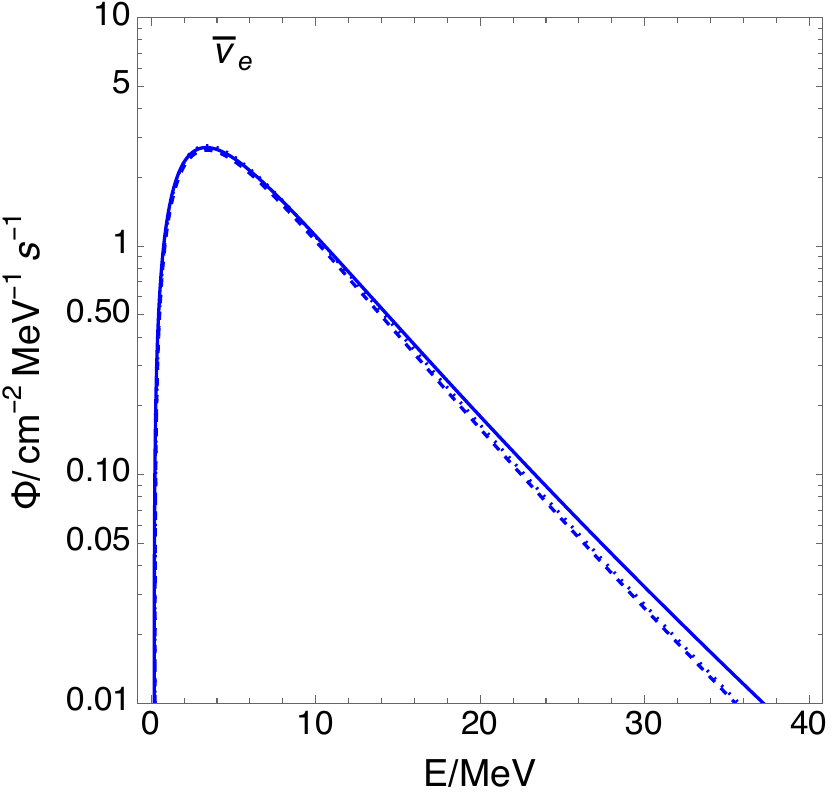}
  \includegraphics[width=0.4\linewidth]{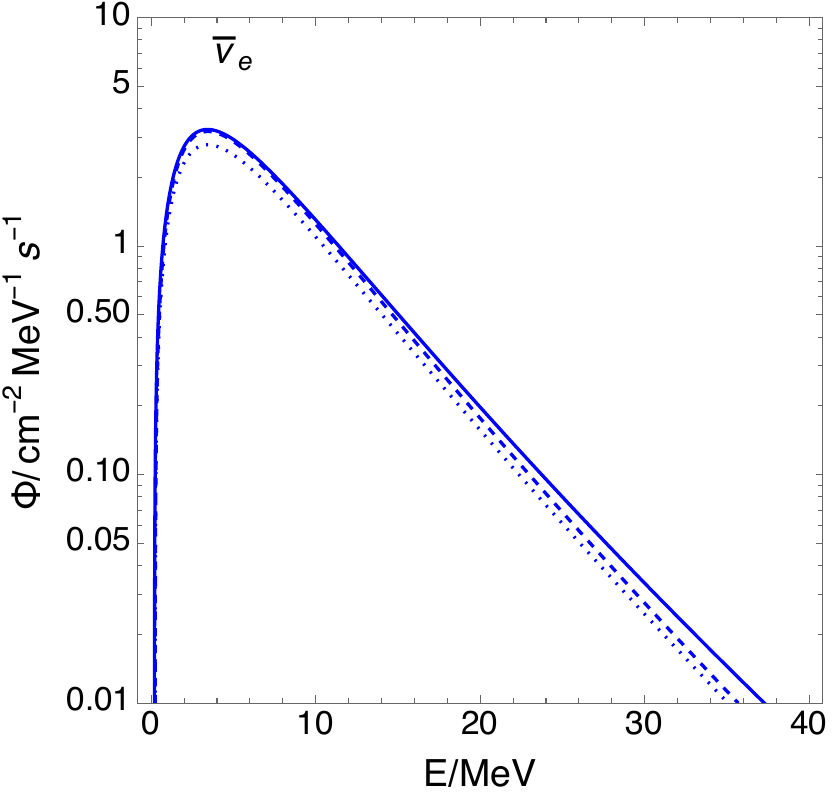}
  \includegraphics[width=0.4\linewidth]{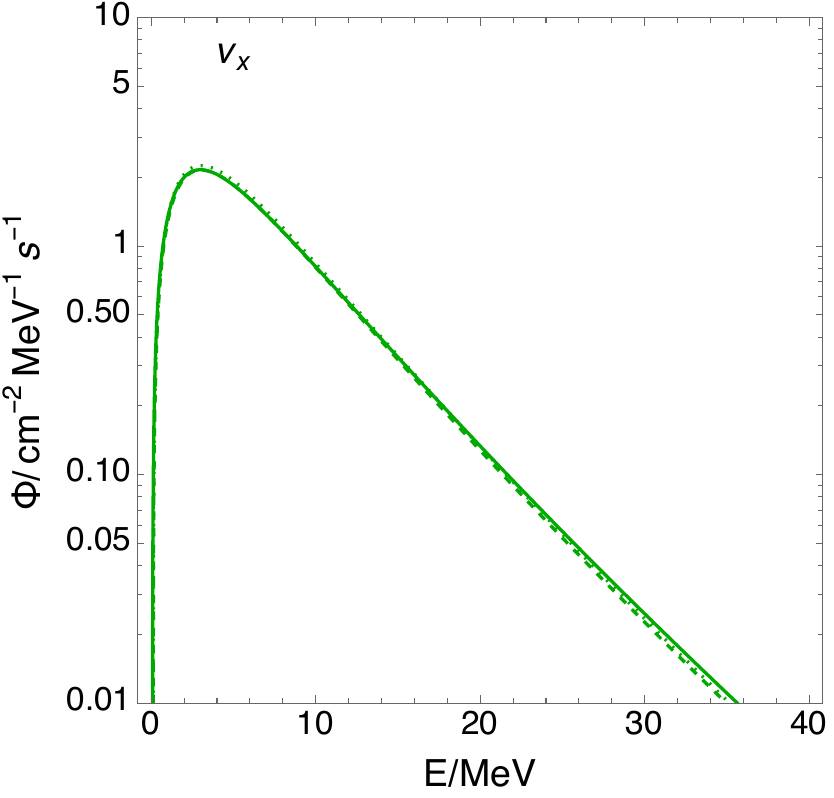}
  \includegraphics[width=0.4\linewidth]{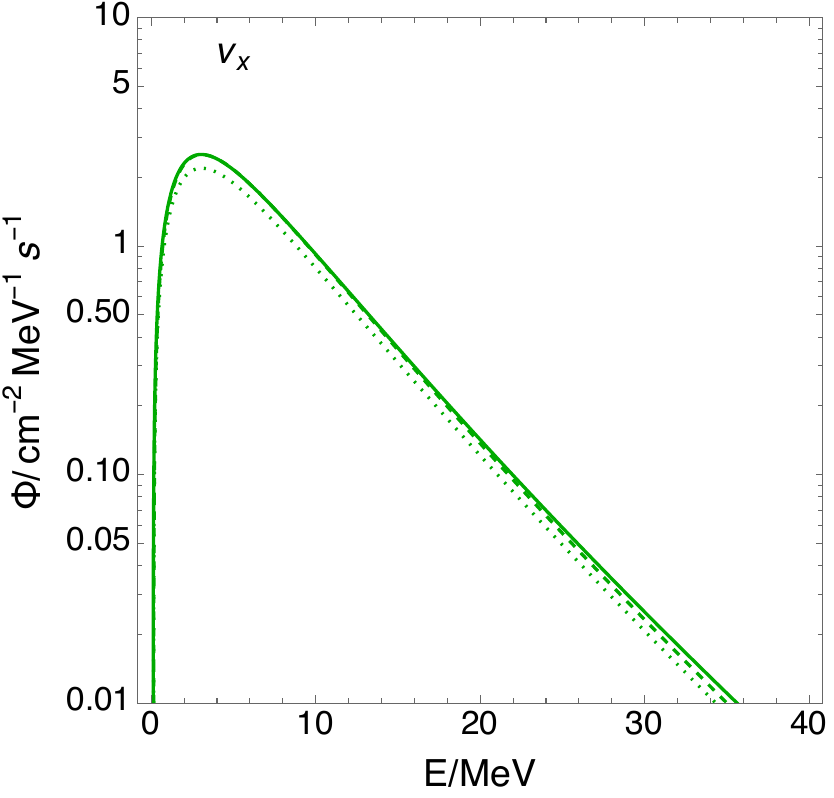}
  \caption{DSNB flux for the different neutrino flavors, for BPASS  the \km\ fiducial model. In the no BH cases, the neutrino spectra for neutron-star-forming collapses (curves in Fig.~\ref{fig:nuparam}) are applied to the entire stellar population.  Flavor oscillations are not included.
  The data is available online, see \cite{Lunardini:2025:DSNB}.
  }
  
  \label{fig:dsnbNoOsc}
\end{figure*}

\begin{figure*}
    \centering
    \includegraphics[width=0.40\linewidth]{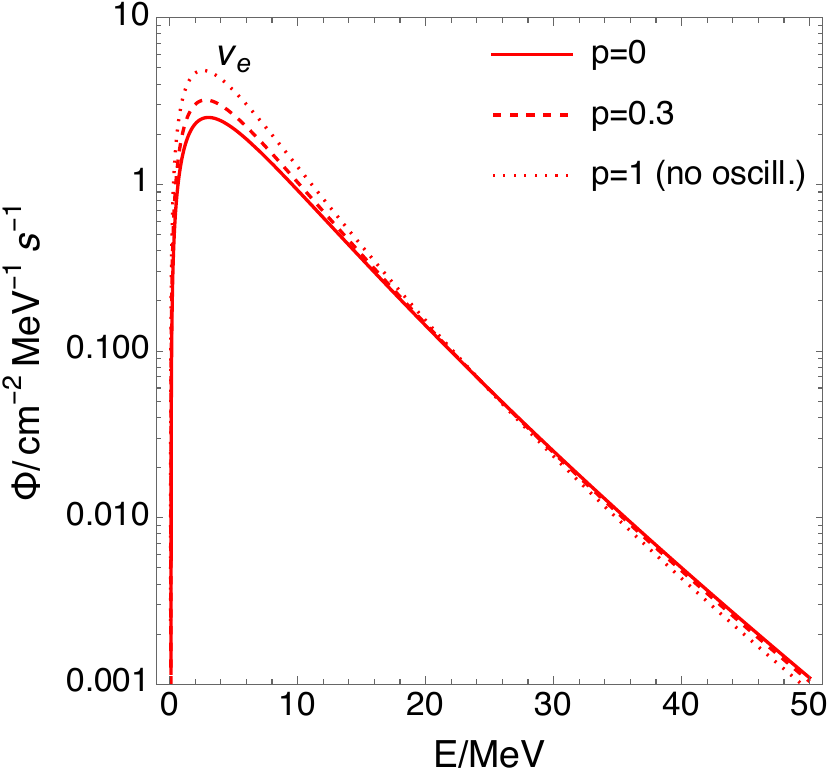}
    \includegraphics[width=0.40\linewidth]{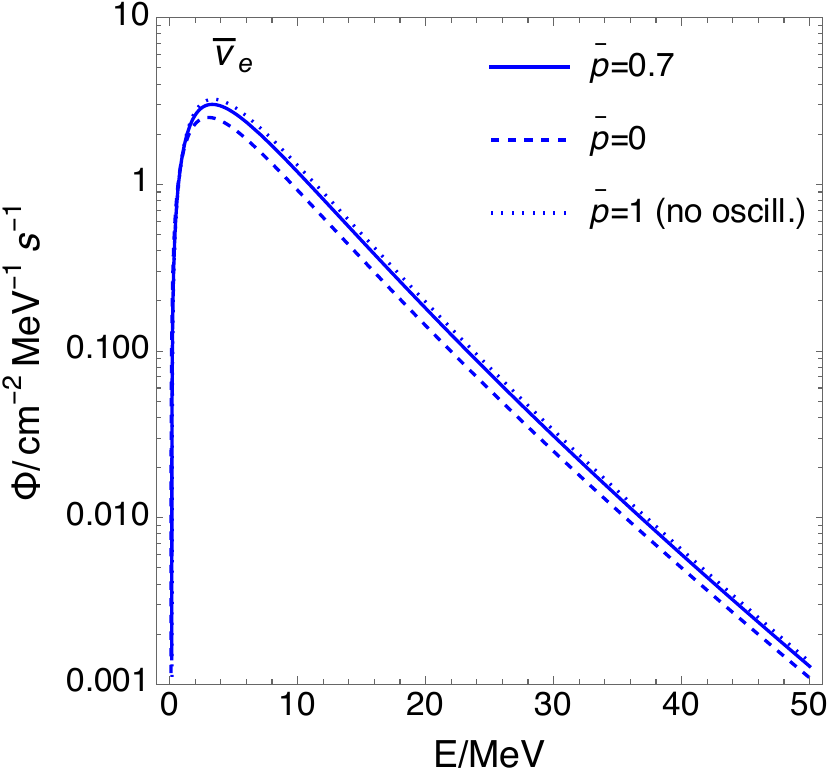}
    \caption{The $\nue$ and $\barnue$ components of the DSNB with the inclusion of BH formation and binary evolution effects.  
    The latter are from the \km\ fiducial model; results for  BPASS are very similar and will not be shown. Here the effect of flavor oscillations is described by the permutation parameters, $p$ and $\bar p$. }
    \label{fig:dsnbMSW}
\end{figure*}

Fig.~\ref{fig:dsnbNoOsc} illustrates the effect of including binary evolution effects and/or black hole formation on the \df, for the un-oscillated flavor fluxes and the two simulations of binary evolution. 

It appears that binary evolution effects are moderate: for the \km\ model, they amount to a $\sim 10$--$15\%$ increase in the normalization of the flux, reflecting the increase in the number of collapse candidates --- especially at low $\mco$ ---  compared to single-star evolution. That is consistent with the results of \citetalias{Horiuchi:2020jnc}, and different from \citet{Kresse:2020nto}, who found that including binary effects {\it decreases} the \df, since mass stripping decreases the CO core mass \cite{Woosley2019ApJ}. 
Mergers of low-mass stars, whose individual masses are below the core-collapse threshold, can increase the DSNB if the combined mass exceeds the threshold. In contrast, mergers between stars already massive enough to undergo core collapse individually tend to reduce the DSNB. The net effect of binary interactions is thus determined by these competing processes~\citepalias[see the discussion in][]{Horiuchi:2020jnc}.

For the case of BPASS, binary evolution effect are at the level of $\sim -4\%$, and are invisible in Fig.~\ref{fig:dsnbNoOsc}.
The difference in the \df\ results between the cases with BPASS and with the \km\ model can partially be traced to the assumptions of the evolution of the core after a stellar merger.  In the \km\ model, the merged remnants are assumed to be highly rotating. The high rotation is then considered to lead to an increase in the core mass during the subsequent evolution.

The effect of including black hole formation leads to a high energy tail in the (un-oscillated) $\barnue$ and $\nue$ spectra, which is consistent with previous literature (e.g., \cite{Lunardini:2009ya}). The flux increase --- with respect to \nsfc\ only  --- is of  $\sim 50\% $ at $E=35$\,MeV. For $\nux$ there is an increase of similar magnitude in the normalization, while the energy spectrum is practically unmodified, due to the very modest difference in $\langle E \rangle_x$ between \nsfc\ and \bhfc\ (Fig.~\ref{fig:nuparam}).

 In Fig.~\ref{fig:dsnbMSW}, the $\barnue$ and $\nue$ components of the \df, $\Phi^{\bar e,i}(E)$ and $\Phi^{e,i}(E)$ (which include flavor oscillations),  are shown, for different values of the flavor-swapping parameters. Binary evolution and black hole formation are included. For both species, the main effect of flavor oscillations is to \emph{decrease} the expected flux, compared to the no-oscillation case. The decrease is largest for $\nue$ near the peak of the flux, $E \sim 3$\,MeV, where it can amount up to a factor of $\sim 2$; it is at the level of $\sim20\%$ or so in the $\barnue$ channel.
 This trend is in contrast with typical results that can be found in the literature, where oscillations tend to increase the $\nue$ and $\barnue$ fluxes, at least in the region of sensitivity of current detectors. The difference is due to the fact that here  the originally produced $\nux$ (part of which is converted into $\nue$ and $\barnue$) have lower luminosity than the other species, and spectra that are very similar to those for $\barnue$. Instead, most literature assume $\nux$ fluxes with (approximately) equal luminosity and harder spectra than the electron flavors.

\section{Summary, discussion and conclusions}
\label{sec:discussion}

We have followed a methodology for modeling the \df\ that is based on the mapping from CO core mass (at advanced pre-supernova stage) to compactness, as summarized in Figs.~\ref{fig:overview} and \ref{fig:introfigure}. The method is suitable for the systematic inclusion of (i) neutrino spectra from large sets of numerical core collapse simulations, and (ii) population synthesis models. We have applied it to state-of-the-art models; in particular,  we have processed the results of the extended suite of simulations from \citet{Vartanyan:2023zlb}, and we have identified the relevant dependence of the \n\ spectra and fluxes on the compactness parameter. We have also analyzed the results of two modern population synthesis codes, BPASS \cite{Patton:2021gwh} and the \km\ model \cite{Kinugawa:2014zha}, and obtained the distribution of the stellar populations with respect to the CO-core mass, with and without binary evolution effects.  

We can now draw a number of conclusions of our work: 

\begin{enumerate}

\item Our analysis of the numerical results of \citet{Vartanyan:2023zlb} for neutron-star-forming collapses shows a dependence of the \n\ flux parameters on $\xic$ which is linear for the total energies emitted (per flavor) and like $\xic^{1/3}$ for the average energies and rms energies. The total energy emitted in each of the non-electron flavors is lower by $\sim 20$--$50\%$ than the electron flavors. For antineutrinos, this is actually the main difference between $\barnue$ and $\nux$, because  the spectra of these two species are very similar. The spectrum of $\nue$ is colder than that of other species (a $\sim 30\%$ difference in the average energies) which consistent with older models. 

\item 
    The impact of \n\ flavor oscillations on the \df\ is smaller than in most previous literature, due to electron and non-electron flavors having similar spectra at the production site, especially in the antineutrino channel. In contrast with the traditional expectation, if the effective survival probabilities (eq.~(\ref{eq:dsnboscill})) are constant, oscillations could mostly modify the flux normalization rather than the energy spectrum, in the energy window of sensitivity of current and near future experiments ($E \gtrsim 10$\,MeV). In the neutrino channel, flavor conversion can result in harder $\nue$ spectra, mainly in the form of a reduction of the flux peak at $E\sim 4 $\,MeV. 
     
\item In the work of \citet{Vartanyan:2023zlb}, black-hole-forming collapses are predicted to occur in $\sim 20\%$ of all collapses, for $2.0<\mco/M_\odot <3.1$ (meaning medium mass progenitors in case of single-evolving stars, $12 \leq M/M_\odot \leq 15.38$). These black-hole forming collapses have more energetic \n\ spectra and higher luminosities, and therefore increase the \df\ in its high energy tail and in its overall normalization. The effect is of $\sim 40\%$--$60\%$ size.

 \item 
 Due to the modest dependence of the \n\ spectral parameters on $\xic$ --- and therefore on $\mco$ ---  the main effect of binary evolution on the \df\ is a moderate change in its normalization. The change is largest, up to a $\sim 15\%$ increase, for the \citetalias{Kinugawa:2014zha} model, and is  caused mostly by the net increase in the number of collapsing stars.
 It is interesting that binary population synthesis  predicts the existence of stars with more massive CO cores, $\mco \gtrsim 10\msun$, for which more energetic \n\ emission (than most single-evolved stars) is expected. However, these high $\mco$ stars amount to only $\sim1\%$ of the population; this fact, combined with weak dependence of the \n\ spectral parameters with $\mco$ results in a negligible effect on the \df.

\end{enumerate}

Before concluding, a number of cautionary remarks are in order. One of them is that our results suffer from uncertainties related to the scatter (see, e.g.,  Fig.~\ref{fig:introfigure}~(b)) and sparseness (e.g., Fig.~\ref{fig:BPASSoverviewKin}, central pane) of the numerical results that we have used. For this reason, we chose to consider small numerical features as not significant, and centered the discussion on then more robust, general trends that we found.

The path and rate of black hole formation remain highly uncertain, and these uncertainties could have a significant impact on the \df. 
In the core collapse simulations we used, a significant fraction of models exhibit long-lived proto neutron stars, $\ge 3.5\,{\rm s}$, without driving the supernova explosion. Typically, however, in the case that the explosion is not initiated, black hole formation occurs within $\mathcal{O} (1\,{\rm s})$ \cite{OConnor:2010moj}, with the timescale depending on the equation of state \cite{Liebendoerfer:2002xn,Sumiyoshi:2007pp,OConnor:2010moj,Nakazato:2013maa,Walk:2019miz}.
In some models, the black hole formation could occur within an exploding model \cite{Chan:2017tdg,Pan:2017tpk,Kuroda:2018gqq,Burrows:2023nlq}. 
Mass accretion  --- enhancing the \n\ emission --- could continue even after the shock revival, with the timescale of $\mathcal{O} (10\,{\rm s})$ and have large influence on DSNB \cite{Nakazato:2024gem,Akaho:2023alv}.
Future studies will require an extensive set of 3D simulations \cite{Bollig:2020phc,Nakamura:2024qhx,Wang:2023vkk} to explore these formation scenarios in great detail.

Some uncertainties are also associated with neutrino flavor conversion. Recently, neutrino oscillations induced by the neutrino-neutrino interactions  -- which could result in energy-dependent flavor survival probabilities --   have been studied in detail \cite{Duan2010,Mirizzi:2015eza,Tamborra2021,Capozzi2022review,Richers2022review,Volpe2023review}. Though self-consistent simulations in several time snapshots have been performed employing this effects \citep{Nagakura:2023xhc,Xiong:2024tac}, we have not found a simple way to incorporate this effect accurately in the global simulations \citep[however see][for the phenomenological modeling]{Ehring:2023abs,Mori:2025cke,Wang:2025nii}.
We have to wait for the establishment of an appropriate treatment of collective oscillations to accurately assess their impact on the \df.

The study of the evolution of binary systems has yet to reach maturity \cite[][]{Tauris:2023nmj,Marchant:2023wno}, and therefore current descriptions have significant uncertainties. It is possible that, as more detailed treatments are developed, a larger impact of binary evolution on the \df\ may emerge. For example, including possible highly-rotating stars may enhance the mixing of compositions and produce larger CO-core masses \cite{Horiuchi:2020jnc}, that will result in higher \n\ emission. Perhaps new explosion channels incorporating rotation and magnetic fields can appear \cite{Summa:2017wxq,Takiwaki:2021dve,Matsumoto:2022hzg,Matsumoto:2023noc,Obergaulinger:2021omt,Shibagaki:2023tmh,Powell:2024nvv,Kuroda:2024xbe}.
Also more sophisticated treatment of stellar mergers could alter the core mass distribution.

In this paper, we did not explicitly model a redshift- or metallicity-dependence of the CO-core mass distribution \((dN/dM_{\rm CO})_i\); however, \(M_{\rm CO}\) may in reality evolve with redshift through metallicity-dependent single- and binary-star evolution and through changes in population properties.  Low-\(Z\) environments favor weaker winds, more efficient rotational/chemically homogeneous evolution, and metallicity-dependent binary evolutions, all of which may  increase CO-core masses \cite[e.g.][]{Belczynski:2010GW-metallicity,Kinugawa:2017LGRB-merger}. A more top-heavy IMF at low \(Z\) is also suggested by recent simulations \cite{Chon:2021jlx,Chon:2024rad-feedback}. Non-universal IMFs were explored in the DSNB context in Ref.~\cite{Ziegler:2022ivq}, but we keep \((dN/dM_{\rm CO})_i\) redshift-independent here for clarity and leave a full \(Z\)- and \(z\)-dependent treatment to future work.

It is important to note that the effects we studied here are comparable or smaller than the uncertainty on the cosmological rate of core collapse, which is several tens of per cent or larger. Therefore, it may not possible to test them in the near future, when the first \df\ data become available. Indirectly, our study contributes to stress the importance of working to reduce the uncertainty on the core collapse rate. 

In closing, we have produced a new, state-of-the-art model for the \df, which includes updated \n\ fluxes and spectra, and effects of binary evolution. We have done so using a modern method that lends itself to further developments, especially those that will be driven by large sets of numerical simulations. Our analysis highlights the potential of the \df\ as a test of a wide variety of physical phenomena --- ranging from the microphysics inside collapsing stars, to the way stars that are born with companions evolve over their lifetime --- and how such potential requires reducing the numerous uncertainties that are present in order to be fulfilled.

\begin{acknowledgments}
We thank the authors of Ref.~\cite{Patton:2021gwh} for sharing their numerical results, and the authors of Ref.~\cite{Vartanyan:2023zlb} for useful discussions and clarifications on their paper. We are also grateful to Hiroki Nagakura, Ko Nakamura, Ken'ichiro Nakazato and Bernhard M\"uller for fruitful discussions. C.L. acknowledges support from the NSF grants 2309973 and 2012195 and from the National Astronomical Observatory of Japan, where this work was started and developed for the most part. She is also grateful to the Institute for Advanced Study of Princeton for the hospitality while part of this work was carried out. The work of SH is supported by NSF Grant No.~PHY-2209420. This work is also supported by JSPS KAKENHI Grant Numbers JP22K03630, JP23H04893, JP23H04899, JP23K20848, JP23K22494, JP23K25895, JP23K03400, JP24K00631 and funding from Fukuoka University (Grant No.GR2302), by MEXT as “Program for Promoting researches on the Supercomputer Fugaku” (Structure and Evolution of the Universe Unraveled by Fusion of Simulation and AI; Grant Number JPMXP1020230406) and JICFuS, and by World Premier International Research Center Initiative (WPI Initiative), MEXT, Japan.

\end{acknowledgments}


\input{draft_dsnb_method.bbl}

\appendix 

\section{Functional fits}

Here we give details on the analytic functions that were found by fitting numerical tables. They should be considered as useful tools, and not necessarily descriptive of the physics underlying the phenomena under consideration. 

\subsection{Compactness as function of CO core mass}
\label{subsub:ximcofun}

We used the function: 
\begin{eqnarray}
\xi_{2.5, {\rm fit}}(\mco)&=&a+b \mco^c \nonumber \\
&+&G_1 e^{-\frac{(\mco-M_1)^2}{\sigma_1^2}}+G_2
   e^{-\frac{(\mco-M_2)^2}{\sigma_2^2}}~,  
   \label{eq:ximcofunfit}
\end{eqnarray}
where masses are in units of $\msun$. 
Here $c$ was fixed at $c=-1$, following Mapelli et al \cite{Mapelli:2019ipt};  this value was found to be close to the actual best fit one. Similarly, $M_1$ and $M_2$ were fixed to convenient values that are found to provide a good fit. All other parameters were fit numerically. Parameter values are are given in Table \ref{tab:xiMcofunfit}.

\begin{table}[htp]
\centering
\caption{Values of the parameters in Eq.~\ref{eq:ximcofunfit}. The values of $M_i$ and $\sigma_i$ are in $\msun$. }
\label{tab:xiMcofunfit}
\begin{tabular}{@{}lc@{}}
\toprule
\textbf{Parameter} & \textbf{Value} \\ 
\midrule
$a$ & 0.39 \\ 
$b$ & -0.76 \\ 
$c$ & -1.0 \\ 
$G_1$ & 0.205 \\ 
$M_1$ & 5.0 \\ 
$\sigma_1$ & 0.382 \\ 
$G_2$ & 0.216 \\ 
$M_2$ & 12.0 \\ 
$\sigma_2$ & 1.22 \\ 
\bottomrule
\end{tabular}
\end{table}

\subsection{Neutrino flux parameters as functions of compactness}
\label{subsub:nuxifun}

For the quantities $E_{{\rm tot},w}$, $\langle E\rangle_w$ and $\langle E_{\rm rms}\rangle_w$ as functions of compactness we adopted the form 
\begin{equation}
f_{{\rm fit}, \nu}=a + b~ (\xic)^c ~. 
\label{eq:nuparamfuncfit}
\end{equation}
Here $a,b,c$ are fitting parameters; their values are given in Table~\ref{tab:nuxifunfit}.

\begin{table}[htbp]
\centering
\caption{The vector $(a,b,c)$ of the fitting parameters in Eq.~\eqref{eq:nuparamfuncfit}, for each neutrino species. $a$ and $b$ are in the appropriate units, namely $10^{52}$ erg when describing $E_{\rm tot}$ and MeV in all other cases.}
\label{tab:nuxifunfit}
\begin{tabular}{@{}lccc@{}}
\toprule
\textbf{Parameter} & \textbf{$\nue$} & \textbf{$\barnue$} & \textbf{$\nux$} \\
\midrule
$E_{\rm tot}$ & $(4.98, 10.5, 0.765)$ & $(4.57, 12.5, 0.809)$ & $(3.73, 12.3, 1.1)$ \\
$\langle E\rangle$ & $(11.5, 2.48, 0.34)$ & $(14.1, 2.78, 0.35)$ & $(13.6, 3.05, 0.34)$ \\
$\langle E_{\rm rms}\rangle$ & $(13.2, 3.95, 0.29)$ & $(16.0, 4.10, 0.30)$ & $(15.8, 4.56, 0.3)$ \\
\bottomrule
\end{tabular}
\end{table}

\end{document}

%% file: draft_dsnb_method.bbl
%